\documentclass[twocolumn,aps,prb]{revtex4-2}
\usepackage[pdftex]{graphicx}
\usepackage{dcolumn}
\usepackage{bm}
\usepackage{float}
\usepackage{hyperref}
\usepackage{epstopdf}
\usepackage{mathrsfs}
\usepackage{graphicx}
\usepackage{multirow}
\usepackage{amsmath}
\usepackage{amssymb}

\newcommand{\ket}[1]{\left| #1 \right>}
\newcommand{\bra}[1]{\left< #1 \right|}

\newcommand{\Fig}[1]{Fig.\,\ref{#1}}
\newcommand{\Eq}[1]{Eq.\,\eqref{#1}}

\hypersetup{
            colorlinks=true,
            linkcolor=blue,
            anchorcolor=blue,
            citecolor=blue}
\begin{document}

\title{Magnon Landau-Zener tunneling and spin current generation by electric field}
\author{YuanDong Wang$^{1,2}$}
\author{Zhen-Gang Zhu$^{1,2}$}
\email{zgzhu@ucas.ac.cn}
\author{Gang Su$^{3}$}
\email{gsu@ucas.ac.cn}
\affiliation{$^{1}$ School of Electronic, Electrical and Communication Engineering, University of Chinese Academy of Sciences, Beijing 100049, China.\\
$^{2}$ School of Physical Sciences, University of Chinese Academy of Sciences, Beijing 100049, China. \\
$^{3}$ Kavli Institute for Theoretical Sciences, University of Chinese Academy of Sciences, Beijing 100190, China.
 }

\begin{abstract}
To control magnon transport in magnetic systems is of great interest in magnonics. Due to the feasibility of electric field, how to generate and manipulate magnon with pure electrical method is one of the most desired goals. We propose that the magnon spin current can be generated by applying a
time-dependent electric field, and the coupling between magnon and electric field is invoked via
Aharonov-Casher effect. In addition, we propose a magnon Landau-Zener tunneling effect that is the key impetus of the induced magnon spin current. Our theory is applied to the 1D ferromagnetic Su-Schrieffer-Heeger model, showing that the generated magnon spin current is closely related to the band geometry. Our findings will expand the horizons of magnonics and electric control-magnonmechanisms.
\end{abstract}
\maketitle


\textit{Introduction.} Magnons, whose quanta are spin-1 spin waves, are the collective excitations of ordered magnets \cite{kittel2005introduction}. Very recently, magnon-based spintronics has attracted much attentions due to distinguishable advantages of magnons.
The charge neutrality, which leads to that the spin current carried by magnons does not incur Joule heating, as well as the  long coherence length of magnons, hold huge potential to realize low-dissipation devices \cite{Kruglyak_2010,kajiwara2010transmission,chumak2015magnon, RevModPhys.90.015005, jungwirth2018multiple}.

The creation and control of magnon spin currents at  nanoscale are key goals in spintronics and magnonics. Much effort has been focused on the thermal control of magnons \cite{PhysRevB.81.214418, PhysRevB.89.014416, PhysRevLett.115.266601, PhysRevLett.116.097204, PhysRevLett.104.066403, onose2010observation, PhysRevLett.106.197202, PhysRevLett.117.217202, PhysRevLett.117.217203, PhysRevB.96.134425, PhysRevResearch.2.013079, du2021nonlinear, PhysRevB.106.035148, PhysRevResearch.4.013186}.
However, the thermal control is not an accurate way in spatial scale and sometimes even cumbersome.
Consequently, a desired goal is to generate and control the magnon transport via electricity.
Nevertheless, owing to the charge neutrality, there does not have a direct coupling between the magnons and electric field.
Recently, increasing interests have been recently devoted to overcoming this difficulty.
For example, it has been proposed that the magnon spin current is generated by circularly polarized light via a two-magnon-Raman process \cite{PhysRevB.104.L100404}.
The optical control of magnons is proposed by making use of the magnetoelectric coupling in multiferroic materials, \cite{PhysRevLett.106.057403, takahashi2012magnetoelectric, bordacs2012chirality}.
Particularly, the dc magnon spin photocurrent has been predicted  in collinear antiferromagnets via the coupling between electric field and polarization with a broken inversion symmetry \cite{PhysRevB.107.064403}.
Noting that the dc magnon spin current is due to the difference of frequencies in response to electric field or magnetic field.
It is regarded as a second-order response, and the signal could be relatively small when compared to the linear response.
Therefore, it is highly desired to find a linear-order means to control magnon response via electric field rather than the second-order response. The reason is that the linear-order response always plays a leading role when it exists.

In this work, we propose such a way by leveraging Aharonov-Casher (AC) effect \cite{PhysRevLett.53.319}, a geometric phase that magnons
pick along the path in the presence of a driving electric field, and to elucidate
it in magnetic insulators.
The effect of driving electric field is described by dipole interaction between the electric field and the magnon position that is  mediated by AC phase.
We solve the time-dependent Sch\"{o}rdinger equation for magnons that does not rely on a nonperturbative expansion on electric field.
We find that the magnon spin current is determined by the transition probability of magnons between different bands, which is associated with the nonadiabatic Landau-Zener tunneling \cite{landau2013quantum, zener1932non, berry1990geometric, joye1991exponential, PhysRevA.61.023402, PhysRevA.66.023404, glasbrenner2023landau, ivakhnenko2023nonadiabatic}.
In contrast to the charge or spin current carried by electrons, the magnon spin current driven by electric field is drastically different in two cases.
Firstly, for  collinear ferromagnet (FM) and antiferromagnet (AFM), the magnitude of the magnon spin current is sourced in the electric field component that is normal to the plane spanned by the direction of magnetization and the magnon movement.
On the other hand, a nonzero magnon spin current requires the electric field to be time-varying; while such constraint is absence for the charge current. Finally we suggest the Su-Schrieffer-Heeger (SSH) FM model as possible candidates to realize the electric field driven magnon spin current.
Our results show that collinear FM and AFM driven by a time-varying electric field can serve as effective spin current generators and provide a promising platform to explore the nontrivial magnonic effects.





\textit{A general description.-}
Applying a time-dependent electric field, a moving magnetic dipole moment (magnon) associated with spin along $z$ acquires a geometric  Aharonov-Casher (AC) phase \cite{PhysRevLett.53.319}:
\begin{equation}\label{ac-phase}
\theta_{ij}=(g\mu_{B}/\hbar c^2)\int_{\bm{r}_i}^{\bm{r}_j}(\bm{E}(t)\times \hat{\bm{e}}_{z})\cdot d\bm{r},
\end{equation}
where $\bm{E}(t)$ is the electric field.
We suppose that the magnetization direction is along $z$-direction, hence the magnetic moment of a magnon is $\bm{\mu} = -g\mu_B \hat{\bm{e}}_z$, with $\mu_B$ the Bohr magneton and $g$ the Land\'{e} factor.
It is noted that the AC phase is a special case of Berry phase \cite{R_Mignani_1991}, in parallel to the well-known AB phase.
For a spin Hamiltonian, it is shown that the electric field can be introduced by  a canonical momentum incorporated with an effective vector potential \cite{PhysRevLett.90.167204, PhysRevB.96.224414, PhysRevB.95.125429}.
Thus, for a collinear ferromagnet, the single-particle Hamiltonian reads \cite{supplementary} 
\begin{equation}\label{hami-velocity}
\mathcal{H}_{A}(\bm{k}, t) = \mathcal{H}_0(\bm{k} - (g\mu_B/\hbar c)\bm{A}_{\text{eff}}(t)),
\end{equation}
where  the effective vector potential is
\begin{equation}\label{a-eff}
\bm{A}_{\text{eff}}(t) = (1/c)\bm{E}(t)\times \hat{\bm{e}}_z.
\end{equation}
\begin{table}[tb]
	\renewcommand\arraystretch{2}
	\caption{List of the analogy between then dipole interaction of magnons via Aharonov-Casher phase and that of electrons via Aharonov-Bohm phase.}\label{tab1}
	\begin{tabular*}{\columnwidth}{p{4cm}<{}p{2cm}p{2.5cm}}
		\hline\hline
         & electron & magnon          \\
		\hline
        ``Charge"
        &$e$ & $g\mu_B /c$
        \\
        \hline
	  Vector potential
	  & $\bm{A}$ & $\bm{A}_{\text{eff}} = \frac{1}{c}\bm{E}\times \hat{\bm{e}}_z$  \\
	  \hline
	  Gauge freedom of $\bm{A}$ and $\bm{A}_{\text{eff}}$
	  & $\checkmark$ & $\times$  \\
        \hline
	  ``Electric field"
	  & $\bm{E}=-\frac{\partial \bm{A}}{\partial t}$ & $\bm{\mathcal{E}} = -\frac{1}{c}\frac{\partial \bm{E}}{\partial t}\times \hat{\bm{e}}_z$  \\
	  \hline
	  Time constraint on $\bm{E}$
	  & $\times$ & $\checkmark$  \\
		\hline\hline
	\end{tabular*}
\end{table}
Alternatively, the perturbed Hamiltonian \Eq{hami-velocity} can be written in form of an effective dipole interaction via a unitary transformation as shown in SM \cite{supplementary} 
\begin{equation}\label{hami-length}
\mathcal{H}_{E}(\bm{k},t)=\mathcal{H}_0\left(\bm{k}\right)+(g\mu_B/ c)\bm{\mathcal{E}}(t)\cdot \bm{r},
\end{equation}
where $\mathcal{H}_0$ is the single-magnon Hamiltonian in crystal momentum space. In \Eq{hami-length} we introduce the effective electric field  via 
\begin{equation}\label{etilde}
\bm{\mathcal{E}}(t)\equiv -\partial \bm{A}_{\text{eff}}(t)/\partial t,
\end{equation}
which is in analogy with the electron dipole interaction using magnetic vector potential.
According to \Eq{a-eff} and \Eq{etilde}, it is seen that the  electric field $\bm{E}$ has to be time-varying, otherwise the effective electric field $\bm{\mathcal{E}}$ as well as the effective dipole interaction become zero (i.e. dc electric field case).
Noting that the perturbation Hamiltonian \Eq{hami-length} of magnons presents a formal \textit{duality} to that of electron system with dipole interaction
perturbation, and consequently an electric field along y-axis breaks the inversion symmetry (x-direction) of the one-dimensional spin chain. However, the effective vector potential $\bm{A}_{\text{eff}}$ does not have a gauge degree of  freedom, in contrast to the magnetic vector potential. The analogy between the electric field perturbed Hamiltonian of magnons and that of electrons are shown in Table. \ref{tab1}.

We recall that the magnon eigenstates in an unperturbed crystal is
$
\mathcal{H}_0\ket{u_n,\bm{k}(t)}=\varepsilon_{n,\bm{k}}\ket{u_n,\bm{k}}
$. When consider the electric field perturbation and make use of the dipole Hamiltonian \Eq{hami-length},
the single-particle time-dependent Sch\"{o}rdinger equation is written as
\begin{equation}
i\hbar \partial_t \ket{\Psi_{m},\bm{k}_0 (t)}=\mathcal{H}_{E}(\bm{k},t)\ket{\Psi_{m},\bm{k}_0 (t)},
\label{td-eq}
\end{equation}
where the general magnon instantaneous state in Bloch representation is given by
\begin{equation}
\ket{\Psi_{m},\bm{k}_0 (t)} = \sum_{n}C_{mn\bm{k}_0}(t)e^{i\gamma_{n\bm{k}_0}(t)}\ket{u_n,\bm{k}_{0}(t)},
\label{snap-psi}
\end{equation}
where  the coefficient $C_{mn\bm{k}_0}(t)$ contains variable $t$ with the subscript $\bm{k}_0$ to keep track of the information that $\bm{k}(t=0)=\bm{k}_0$, and the subscript $m$ in $\ket{\Psi_{m\bm{k}_0 }(t)}$ denotes the  initial condition $C_{mm\bm{k}_0}(t_0)=1$.  The phase $\gamma$ including a dynamical and a geometric phase is
\begin{equation}\label{geo-phase}
\gamma_{n\bm{k}_0} =\frac{1}{\hbar}\int_{0}^{t}dt_1 \left[\varepsilon_n(\bm{k}_0 (t_1))  +\frac{g\mu_B}{c}\bm{\mathcal{E}}(t_1)\cdot \bm{\mathcal{A}}_{nn}(\bm{k}_0(t_1)) \right]
\end{equation}
in which $\bm{\mathcal{A}}_{nn}=i\bra{u_n,\bm{k}(t)}\nabla_{\bm{k}}\ket{u_n,\bm{k}(t)}$ denotes the intraband Berry connection.  Recently the nonlinear response has been investigated at great length, and it has been discovered that the geometric nature of the wave functions plays a significant role in the Landau-Zener tunneling \cite{PhysRevB.102.245141, kitamura2020nonreciprocal, 10.21468/SciPostPhys.11.4.075}.
In band insulators with PT symmetry, it has been revealed that the spin current carried by electrons can be generated, which is raised by Landau-Zener tunneling in strong dc electric field \cite{PhysRevB.105.075201}.
Inspired by the progress in electron systems, we propose the tunneling concept to magnon transport.
By inserting \Eq{snap-psi} and \Eq{geo-phase} into  
\Eq{td-eq} and making use of the initial conditions, we obtain \cite{supplementary} 
\begin{eqnarray}
i\partial_t C_{ln\bm{k}_0}(t) &=& \frac{g\mu_B}{\hbar c}\sum_{m\neq n}{\mathcal{E}^\alpha}\lvert \mathcal{A}^{\alpha}_{nm}(\bm{k}_0(t))\rvert e^{i\arg \mathcal{A}^{\alpha}_{nm}(\bm{k}_0(t_0))}  \notag\\
&\times & e^{ i\int_{t_0}^{t_1}dt_2 \Delta_{mn}(\bm{k}_0(t_2))/\hbar}C_{lm\bm{k}_0}(t),
\label{cln}
\end{eqnarray}
where $\mathcal{A}_{nm}^\alpha$ is the inter-band Berry connection and $\Delta_{nm}^\alpha(\bm{k})$ is introduced as
\begin{equation}\label{delt}
\Delta_{nm}^\alpha(\bm{k}) = \varepsilon_{n}(\bm{k})-\varepsilon_{m}(\bm{k})  +\frac{g\mu_B}{c}\mathcal{E}^\alpha (t) {R}_{nm}^{\alpha\beta}(\bm{k}),
\end{equation}
where ${R}_{nm}^{\alpha\beta}(\bm{k}) = {\mathcal{A}}_{nn}^\beta(\bm{k})-{\mathcal{A}}_{mm}^\beta(\bm{k})-  \partial_{k_\beta} \arg \mathcal{A}_{nm}^\alpha(\bm{k})$ is the magnon shift vector \cite{PhysRevB.107.064403, wang2023magnon}, which characterizes the spin polarization difference between two magnon bands $m$ and $n$, which shares a similar expression with the electronic shift vector in semiconductors \cite{PhysRevB.61.5337, PhysRevLett.109.116601, cook2017design, kitamura2020nonreciprocal}.
According to \Eq{hami-velocity},  the gauge-invariant crystal momentum is $\bm{k}=\bm{k}_0 - {(g\mu_B/\hbar c)\bm{A}_{\text{eff}}(t)}$, and the eigenstates of the time-dependent Hamiltonian can be labeled by a single parameter $\bm{k}$.
Noting that $\bm{\mathcal{E}}(t) = -\partial \bm{A}_{\text{eff}}/\partial t$,  the equation of motion of $\bm{k}$ is given by
\begin{equation}\label{k-eq}
\hbar\dot{\bm{k}}= -(g\mu_B/c)\bm{\mathcal{E}}(t).
\end{equation}
By integrating \Eq{cln} from $t_0$ to $t$, we have
\begin{widetext}
\begin{equation}\label{cnm-matrix}
\begin{aligned}
&\begin{pmatrix}
C_{nm\bm{k}_0 (t)} e^{-i\arg \mathcal{A}_{mn}^\alpha(\bm{k}_0)}\\ C_{nn\bm{k}_0 (t)} e^{i\arg \mathcal{A}_{mn}^\alpha(\bm{k}_0)}
\end{pmatrix}=\exp \left[-i\int_{t_0}^{t}dt_1 \frac{g\mu_B}{\hbar c}\mathcal{E}^\alpha(t_1)\vert \mathcal{A}_{mn}^\alpha(\bm{k}_0(t_1))\vert \begin{pmatrix}
0 & W(t_1)  \\ W^*(t_1) & 0
\end{pmatrix} \right]\begin{pmatrix}
0\\1
\end{pmatrix}.
\end{aligned}
\end{equation}
\end{widetext}
where $W(t_1)=\exp \left[i\int_{t_0}^{t_1}dt_2 \Delta_{mn}(\bm{k}_0(t_2))/\hbar\right] $.  In deriving \Eq{cnm-matrix} we used \Eq{k-eq} and the initial conditions  $
C_{mm\bm{k}_0}(t_0)=1
$.
The  probability of magnons with momentum $\bm{k}_0$ tunneling from $n$-band to $m$-band is
\begin{equation}
P_{nm\bm{k}_0}(t) = \vert\langle m ,\bm{k}_0(t)\ket{\Psi_n,\bm{k}_0(t)}\vert^2 = \vert C_{nm\bm{k}_0}(t)\vert^2.
\end{equation}
Noting that $P_{nm\bm{k}_0}(t)$ is gauge-invariant because of the gauge-invariance of $\mathcal{A}_{mn}^{\alpha}$ and ${R}_{nm}^{\alpha\beta}$.
The $z$-direction polarized magnon spin current is given as an expectation value of the magnon spin current operator for all magnons in BZ, which is given by
\begin{equation}
\bm{J}^{s}_z =\sum_{\bm{k}_0\in BZ}\sum_{n}\hbar f_{n,\bm{k}_0}\bra{\Psi_{n},\bm{k}_0(t)} \hat{\bm{J}}^{s}_{\bm{k}_0, z}(t) \ket{\Psi_{n},\bm{k}_0(t)},
\label{js-main}
\end{equation}
where $f_{n,\bm{k}_0}=1/(e^{-\beta\varepsilon_{n}(\bm{k}_0)} -1)$ is the Bose-Einstein distribution with the chemical potential set to zero (since the magnon number is not conserved), $\hat{\bm{J}}^{s}_{\bm{k},z}$ is the $z$-direction polarized magnon spin current operator in $\bm{k}$ space, which is defined  via the  continuity equation (in the long-wavelength limit $\bm{q}\rightarrow 0$) of the local magnon density operator $\hat{n}_{z}(\bm{r}_i)=\hbar \sum_m a_{i,m}^\dagger a_{i,m}$: $\partial \hat{n}_{z\bm{q}}/\partial t + i\bm{q}\cdot \hat{\bm{j}}_{z}^s = 0$.  By use of the periodic boundary conditions, \Eq{js-main} can be alternatively written as \cite{supplementary} 
\begin{equation}
\bm{J}^{s}_z= -\sum_{n,m}\int \frac{d\bm{k}_0}{2\pi} f_{n,\bm{k}_0}\nabla_{\bm{k}}\left[\varepsilon_{nm}(\bm{k})P_{mn}(\bm{k})\right]\vert_{\bm{k}=\bm{k}_0(t)},
\label{js-main2}
\end{equation}
where $\varepsilon_{nm}$ is introduced as $\varepsilon_{nm} = \varepsilon_n - \varepsilon_m$. In \Eq{js-main2}  we use the identity $C_{mn} = C_{nm}^{*}$.



\begin{figure}[tb]
\centering
\includegraphics [width=2.2in]{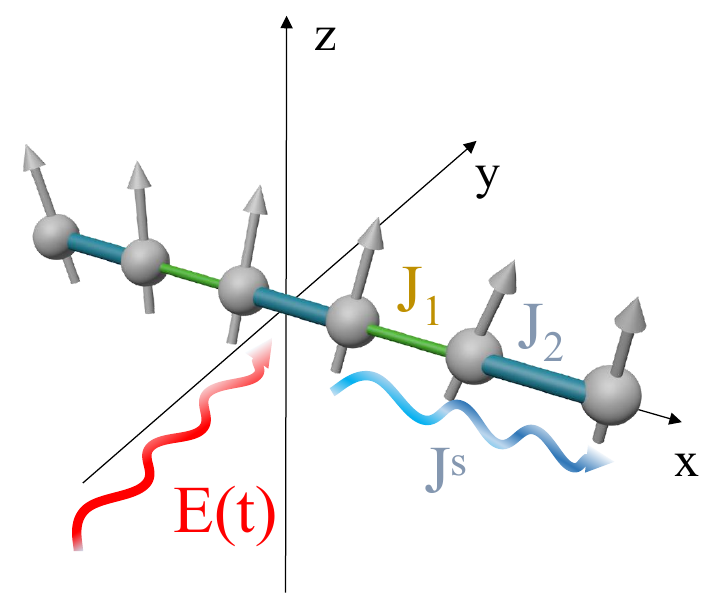}
\caption{Magnon spin current generated by ac electric field for a 1D FM SSH model. The ac electric field with red arrow is perpendicular to the spin chain and the magnetization axis.}\label{fig1}
\end{figure}

\textit{Application to 1D ferromagnetic SSH model.-}
To elucidate the emergence of electric-field-driven magnon spin current, we apply our theory to 1D ferromagnetic (FM) SSH chain with Hamiltonian \cite{Wei_2022}
\begin{equation}\label{hami1}
H = -J_1 \sum_{i=1}^{N}\bm{S}_{i,A} \cdot \bm{S}_{i,B}-J_2 \sum_{i=1}^{N-1}\bm{S}_{i,B} \cdot \bm{S}_{i+1,A}-\sum_{i}K_{i}(S_{i}^z)^2,
\end{equation}
where $J_1$ and $J_2$ represent the exchange interaction in one unit cell (intracellular) and between two unit cells (intercellular), respectively.
The last term is the easy-axial anisotropy term for the quantization $z$-axis, where $K_i$ is the axial anisotropy energy.
From an inspection of the cross product form of \Eq{ac-phase}, for nonzero magnon spin current, a requirement is that the electric field should have nonzero component perpendicular to the magnetization direction.
Another requirement from the direct product form in  \Eq{hami-length} is that the electric field should have nonzero component perpendicular to the 1D spin chain.
As shown in \Fig{fig1},  the 1D spin chain and the generated magnon spin current is along $x$-axis with the magnetization along $z$-axis, and the time-varying electric-field is applied in $y$-direction.

The linear Holstein-Primakoff (HP) transformation gives the magnon Hamiltonian with bosonic generation (annihilation) operator (for $A$ sublattice, $a_{A}^\dagger$ ($a_{A}$); for $B$ sublattice, $a_{B}^\dagger$ ($a_{B}$)) reads 
\begin{equation}
H =H_1+H_2+H_3, 
\notag
\end{equation}
where $H_1=(J_1+J_2 +K) S\sum_{i=1}^{N}(a_{i,A}^\dagger a_{i,A} + a_{i,B}^\dagger a_{i,B})$, $H_2=-J_1 S\sum_{i=1}^{N}(a_{i,A}^\dagger a_{i,B} + a_{i,B}^\dagger a_{i,A})$, and $H_3=-J_2 S\sum_{i=1}^{N-1}(a_{i,A}^\dagger a_{i+1,B} + a_{i+1,B}^\dagger a_{i,A})$. In k-space it is given as $H = \Phi_{k}^{\dagger}\mathcal{H}_{k}\Phi_{k}$, where $\Phi^{\dagger} = (a_{k,A}^{\dagger},a_{k,B}^{\dagger})$, and the single-particle Hamiltonian is
\begin{equation}\label{sp-hami-mag}
\mathcal{H} (k)=\varepsilon_0 I + \Re f(k) \sigma_x + \Im f(k)\sigma_y,
\end{equation}
in which $\varepsilon_0 = (J_1 + J_2 + K)S$, $f(k) = -J_1 S -J_2 S e^{-ika}$ with $a$ being the lattice constant, and $I$, $\sigma_x$, $\sigma_y$ being the unit and Pauli matrices.
%
%
According to \Eq{js-main2}, for 1D two bands system, the expression of magnon spin current reduces to \begin{equation}\label{js-twoband}
{J}^{s}_z= \int \frac{d{k}_0}{2\pi} f_{+-,{k}_0}\partial_{{k}}\left[\varepsilon_{+-}({k})P_{-+}({k})\right]\vert_{k=k_0(t)},
\end{equation}
where we introduce $f_{+-}=f_{-} - f_{+}$.

\begin{figure}[tb]
\centering
\includegraphics [width=\columnwidth]{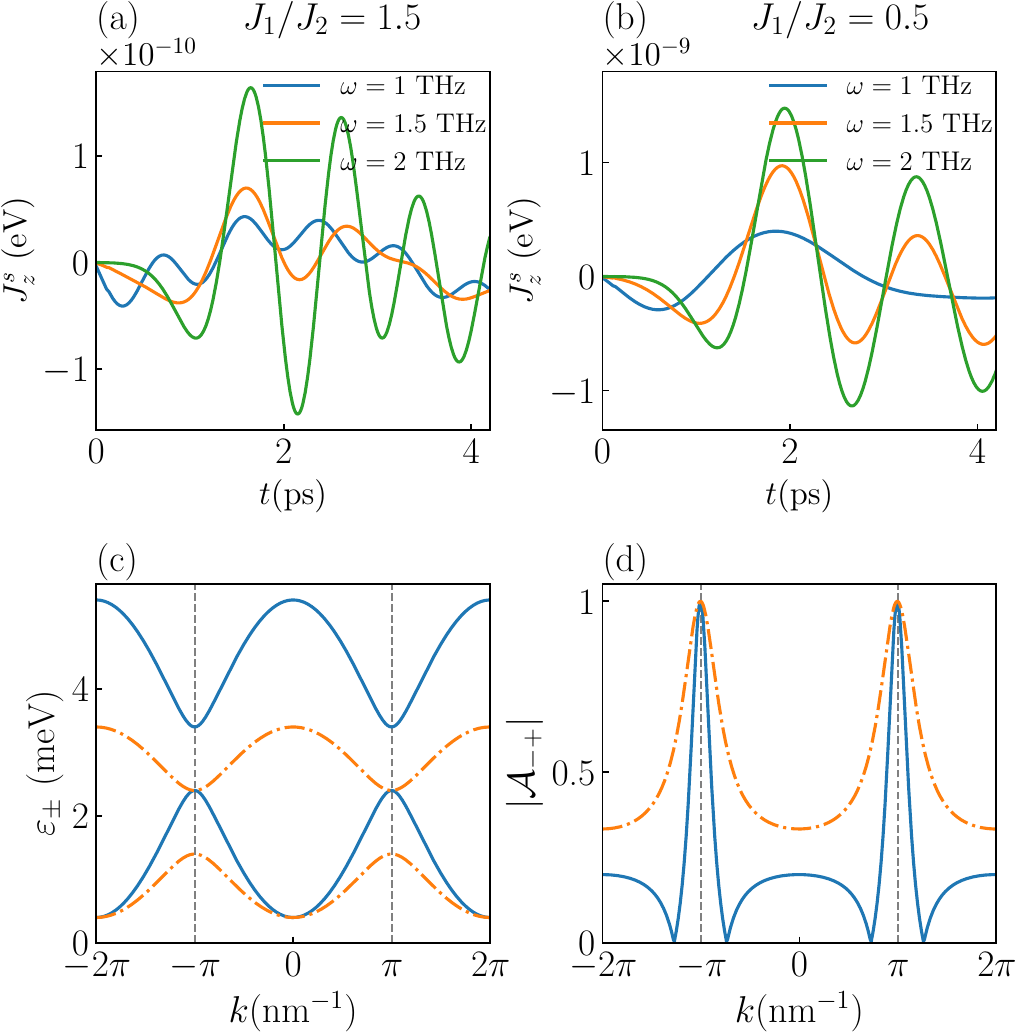}
\caption{Magnon Landau-Zener tunneling in 1D FM SSH model with both effective TRS and inversion symmetry.  Time-evolution of magnon spin current by applying Gaussian pulse for (a) topological trivial phase $J_1/J_2=1.5$  and (b) topological nontrivial phase $J_1/J_2=0.5$ . (c) The band dispersion. (d) The absolute value of the interband Berry connection $\mathcal{A}_{-+}$. The solid (dashed-dot) line corresponds to the ratio $J_1/J_2=1.5$ ($J_1/J_2 = 0.5$).  Parameters are $J_2=1$meV, $K=0.4$meV, $k_B T = 1$meV, $a=1$nm, $E_0 = 2\times 10^{5}$ V/m, $t_0 = 1.5$ ps.}\label{fig2}
\end{figure}

Now we give a symmetry analysis.
The 1D FM SSH model preserves the spatial inversion symmetry, which is
$\mathcal{P}\mathcal{H}(k)\mathcal{P}^{-1} = \mathcal{H}(-k)$,
with the inversion operation $\mathcal{P}=\sigma_x$.
Under $\mathcal{P}$, the Berry connections satisfy $\mathcal{A}_{mn}^\alpha(k)=-\mathcal{A}_{mn}^\alpha(-k)$, and the shift vector transforms as $R_{mn}^{\alpha\beta}(k)=-R_{mn}^{\alpha\beta}(-k)$.
However, the  time-reversal symmetry (TRS) $\mathcal{T}$  is not preserved. Instead, it respects to the effective TRS $\mathcal{T}^\prime$, which is a combination of TRS $\mathcal{T}$ and a spin rotation by $\pi$ about the direction  perpendicular to the quantization axis, i.e.,
$\mathcal{T}^\prime\mathcal{H}(k)(\mathcal{T}^{\prime})^{-1} = \mathcal{H}(-k)$,
which gives a constraint of Berry connections $\mathcal{A}_{mn}^\alpha(k)=\mathcal{A}_{nm}^\alpha(-k)$, and the shift vector  $R_{mn}^{\alpha\beta}(k)=R_{mn}^{\alpha\beta}(-k)$.
Combining the inversion symmetry and the effective TRS, it yields $R_{mn}^{\alpha\beta}(k)=0$. As we will discuss below, nontrivial effect is associated with finite $R_{mn}^{\alpha\beta}$ owing to broken inversion symmetry.

\textit{Electric field driven magnon spin current.-}
For a 1D infinite system, the topological properties in the momentum k-space are often described by the Zak phase $\varphi_{\text{Zak}}(n) = \int_{\text{BZ}}\mathcal{A}_{nn} dk$.   For the magnon Hamiltonian \Eq{sp-hami-mag}, a topological phase transition occurs at a critical ratio $J_1/J_2=1$ accompanied by  energy gap closing \cite{Wei_2022}, which distinguishes the topological trivial phase ($J_1/J_2>1$) and topological nontrivial phase ($J_1/J_2<1$).

Unlike its magnetic field companion the Aharonov-Bohm phase, the coupling between the magnetic moments and electric field via AC phase is generally small owing to the factor $1/c$.  However, as indicated in \Eq{geo-phase}, the geometric phase  scales as the effective field ${\mathcal{E}(t)}$, which means it can be enhanced by increasing the  frequency or the time-derivative of electric field. Let us considering a Gaussian pulse $E(t) = E_0 \exp[-\omega^2(t-t_0)^2 ]$, with $1/\omega$ being half-width of half-maximum, $t_0$ being the center of the pulse.

The band dispersion $\varepsilon_{\pm}$ and the absolute value of interband Berry connection $\lvert\mathcal{A}_{-+}\rvert$ are shown in \Fig{fig2}(c) and \Fig{fig2}(d). The gap locates at $k=\pm \pi$, where $\lvert\mathcal{A}_{-+}\rvert$ peaks. 
The time evolution of the magnon spin currents driven by the Gaussian pulse for different $\omega$ is depicted in \Fig{fig2}(a) for topological trivial phase $J_1/J_2=1.5$ and \Fig{fig2}(b) for topological nontrivial phase $J_1/J_2=0.5$. 
The overall characteristic for both cases is that the currents oscillates with time, and  the maximum amplitude of the oscillation is determined by the center of the pulse,  around $t_0 =1.5$ ps. 
The oscillation manifests itself as a typical evidence of Landau-Zener process, owing to the fact that  the Hamiltonian $H(k(t))$ is time-periodic with period $T=2\pi \hbar c/(g\mu_B \mathcal{E} a)$, which is in analogy with the  Bloch oscillation of the electrons \cite{PhysRevB.105.075201}.  
It is worth noting that the oscillation amplitudes increase with the rising of $\omega$ for both cases, which is consistent with the theory. Moreover, it is found that the magnon spin current for the topological nontrivial phase is an order of magnitude larger than that of the topological trivial phase, it can be understood from behavior of $\lvert\mathcal{A}_{-+}\rvert$. 
As shown in \Fig{fig2}(d), despite that $\lvert\mathcal{A}_{-+}\rvert$ for both topological trivial and nontrivial phase approaches 1 at $k = \pm\pi$, the integral of $\lvert\mathcal{A}_{-+}\rvert$ of $k$ for the topological nontrivial phase is much larger than that of the trivial phase. 
According to  \Eq{cnm-matrix}, it renders larger inter-band transition probabilities for topological nontrivial phase and hence the larger current. We consider the energy scale of the spin chain model of meV and $a\sim$ nm to estimate the magnon spin current. 
The required magnitudes of parameters of electric field to produce a magnon spin current of $J_{z}^s \sim 10^{-10}$ eV are $E_0  \sim 10^5$ V/m, $\omega \sim$ THz, which are accessible in experiments. 
Considering a bundle of the spin chains with distance $a$, it is found that the spin current density to be $J_{z}^s \sim 10^{-17}$ J/$\text{cm}^2$, which is observable based on a Boltzmann theory calculation for spin Seebeck effect in a ferromagnet \cite{PhysRevLett.122.197702,hirobe2017one}.

\begin{figure}[htbp]
\centering
\includegraphics [width=\columnwidth]{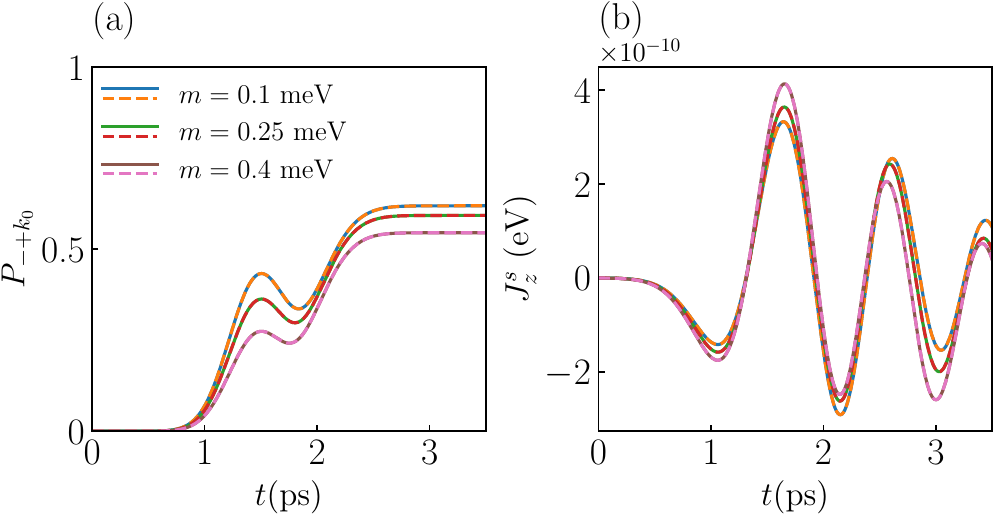}
\caption{Magnon Landau-Zener tunneling in 1D FM SSH model with broken inversion symmetry.  (a) Tunneling probability $P_{-+k_0}$ for wave vector $k_0 = 0$ as a function of time with opposite direction of electric field.   (b) The time-evolution of magnon spin current with opposite direction of electric field. The solid (dashed-dot) line corresponds to $E_0=4\times 10^5$ V/m ($E_0=-4\times 10^5$ V/nm).  The on-site energy $m=0.1$ meV, other parameters are same as that in \Fig{fig2}. }\label{fig3}
\end{figure}

\textit{Reciprocal magnon transport.-}
Noting that for Hamiltonian \Eq{hami1} the magnon spin current is reciprocal owing to the inversion symmetry. When the inversion symmetry is broken, the nonreciprocal transport behaviour is expected. 
The nonreciprocal phenomena of electron transport in noncentrosymmetric system have been extensively studied \cite{PhysRevLett.94.016601, pop2014electrical, doi:10.1126/sciadv.1602390}, and recently it has been shown that the electron transport nonreciprocity is associated with Landau-Zener tunneling  \cite{kitamura2020nonreciprocal}.  
In a noncentrosymmetric magnetic-ordered state, such nonreciprocity has  also been  reported for magnons \cite{PhysRevB.92.184419, PhysRevLett.119.047201, doi:10.7566/JPSJ.88.081007}.

Now we investigate the possible magnon spin current nonreciprocity in the 1D FM SSH model. We add an onsite energy for the $A$ and $B$ sites of $m$ and $-m$ respectively, that modifies the Hamiltonian by $H_{m} = m\sum_{i=1}^{N}(a_{i}^\dagger a_i - b_{i}^\dagger b_i)$ and breaks the inversion symmetry. 
Consequently, it adds a $\sigma_z$ term on the single-particle Hamiltonian, which is written as $\mathcal{H}_m = m \sigma_z$. 
In \Fig{fig3}(a) we depict the time-evolution of the tunneling probabilities $P_{-+k_0}$ (as an example, for the  topological trivial phase) of  wave vector $k_0 = 0$ by applying opposite electric field. 
It shows that the two lines of $P_{-+k_0}$ are overlapping  despite the opposite direction of electric field, yielding a reciprocal magnon spin current as shown in \Fig{fig3}(b) (noting that in \Fig{fig3}(b) we plot $J_{z}^{s}(E)$ and $-J_{z}^s(-E)$ to explicitly show the overlapping).  
However, as mentioned, the broken  inversion symmetry leads to finite shift vector $R_{mn}(k)$, which yields nonreciprocal current \cite{kitamura2020nonreciprocal}.
According to \Eq{delt}, the contribution of shift vector to $\Delta_{nm}$ scales as $1/c$.  For $E_0\sim 10^{5}$ V/m, $\omega\sim$ THz and $a \sim$ nm, it results that $\frac{g\mu_B}{c}\mathcal{E}R_{nm}\sim 10^{-3}$ meV,  which is negligible compared to the energy scale meV of band dispersion. It explains the absence of nonreciprocal current even with broken inversion symmetry.

Finally, several comments are listed in orders. Firstly, noting that the lifetime of magnons is generally finite, which can be captured by the Landau-Lifshitz-Gilbert (LLG) equation 
$\frac{d\bm{M}}{dt} = \gamma \mu_0 \bm{M}\times \bm{H}_{\text{eff}} +\frac{\alpha}{M_S}\bm{M}\times \frac{d\bm{M}}{dt}$, 
with $\bm{H}_{\text{eff}}$  the  effective internal magnetic field, $\gamma$ the absolute value of the gyromagnetic ratio, $\mu_0$ the vacuum permeability, and $\alpha$ the Gilbert damping constant. 
The characteristic time of magnon relaxation  is related to $\alpha$, which is given by \cite{stancil2009spin} 
$\tau = \left( \alpha \omega_k \frac{\partial \omega_k}{\partial \omega_0} \right)^{-1}$, with $\omega_0 = \gamma \mu_0 H_{\text{eff}}$, $\omega_k$ is the spin-wave dispersion. %
Accordingly, it is estimated that $\tau$ is on the order of ns in metallic ferromagnets. In our model, the characteristic time of magnon dynamics is on the order of ps (which is required by the frequency condition), yielding that the damping effect of magnons can be safely neglected. 
Secondly, according to the theory,  fast varying electric fields are able to generate an observable magnon spin current, for another example, a sinusoidal electric field (as shown in SM \cite{supplementary}).
Thirdly, other magnetic states  such as antiferromagnetic, spiral, chiral states  potentially have different advantages such as faster response \cite{nvemec2018antiferromagnetic}, it is worthy pointing out that our theory can be directly generalized to more complicated magnetic states, which deserves a further study.

\textit{Acknowledgments.-} This work is supported in part by National Key R\&D Program of China (Grant No. 2022YFA1402802),  NSFC (Grants No. 11974348, No. 12404147, and No. 11834014), the Strategic Priority Research Program of CAS (Grants No. XDB28000000, and No. XDB33000000), the Training Program of Major Research plan of
the National Natural Science Foundation of China (Grant No. 92165105), and CAS Project for Young Scientists in Basic Research Grant No. YSBR-057.


%

\end{document}